\documentstyle[12pt]{article}
\begin{document}

\author{Y.N. Srivastava, A. Widom and E. Sassaroli  \\
Physics Department, Northeastern University, Boston U.S.A\\ and\\ Physics
Department \& INFN, University of Perugia, Perugia, Italy}
\title{Real and Virtual Strange Processes}
\maketitle

\begin{abstract}
Following notions of quantum mechanics as interpreted by the
Copenhagen School, we make a distinction between measurements involving
one or two virtual $K$ mesons. New predictions result for the
period of $K$ oscillations at the $\Phi$ factory.
\end{abstract}

\section{Introduction}

Our purpose is to discuss the insight that may be gained from $\Phi $
factory experiments $(\Phi \rightarrow K^o+\bar{K}^o)$ concerning a central
issue of quantum measurements; i.e. {\it What is virtual and what is real?}
Before discussing our concrete predictions (i.e. numbers!) for the outcome
of some $\Phi $ factory experiments, we wish first to discuss qualitatively
that {\it not all} aspects of quantum measurements have simple agreed upon
answers. Briefly, we have a great deal of sympathy with Einstein who lost
the debate with Bohr. Known experiments do fit into Bohr's scheme and
mere ``thought experiments'' get boring after a while. Einstein and Bohr did
agree upon what the quantum mechanical scheme entailed, but disagreed on
whether this scheme was the complete story \cite{s1,s2,s3}.

A primary rule of the scheme is that all experimental data are classical.
Suppose that experimental data are stored on (say) a computer disk. Now
imagine that the magnetic grains on the disk were in superposition (an
amplitude for ``0'' and an amplitude for ``1'' on all of the little magnetic
bits in some big binary file). If the computer sends to the standard output
the message ``disk unreadable'', then you know you have no data. Real data
are ``classical'' to a sufficient degree of accuracy. (A few coherent quantum
bits here or there might  appear in an ``error bar'', but if you are
really in quantum superposition, you have no data.) We call data {\it real}.
The primary rule is that data are classical. The classical part of physics
is essential to quantum mechanics in the Bohr scheme. Superposition of
amplitudes applies to those parts of the physical system that you {\it do
not see}. There is no data on a quantum object interacting with the
classical measurement apparatus. If you have data, then it is from the
classical part of the physical system. What you do not see, i.e. the quantum
part of the physical system, is what we call {\it virtual}. The apparatus
is real, and the quantum object is virtual. According to the dictates of
Bohr, a measurement is an interaction between a classical apparatus and a
quantum object. That was decided by Bohr, Heisenberg, and other assorted
friends of the Copenhagen School. That is the conventional scheme of things.
We have reviewed all this because over the years many workers have distorted
what others previously agreed constituted the quantum mechanical analysis of
measurements. For example, we offer a \$100.00 reward to the first reader
who finds for us a manuscript with Bohr as an author which includes the
so-called ``collapse of the wave function'' as a part of quantum mechanical
theory.

Briefly, in the canonical ``electron moves through two slits thought
experiment'' you get quantum interference when there are no data on which
slit the electron passed through. The electron path is then ``virtual''. If
data exist showing which slit the electron chooses, then electron is real
and there is no interference. The electron is real when you see it and
virtual when you do not see it.

Bohr found no problems. The outcome of an experiment depends on how you set
up the real (classical) apparatus, and PLEASE do not ask for data from a
virtual (quantum) object. JUST DO NOT ASK WHAT THE QUANTUM OBJECT IS DOING!
If you get the data, then the object is {\it not quantum}. Do not ask
because there will be no data! Einstein had some problems. He asked
questions. Einstein looked at the moon and it was real. Einstein stopped
looking. Then what? We think the moon is still real, but the contrary has
been proposed by some perfectly competent physicists. So it is our {\it
prejudice} that the moon is real! And we really do not care who looks at it.

\section{What is Virtual and What is Real in Particle Physics?}

In particle physics there are theoretical and experimental views of the
matter. From the theoretical view we would take as ``virtual'' the somewhat
technical but conventional definition of those processes in the ``internal
parts'' of the Feynman diagrams. Consider the Feynman diagram in Fig.1.
 The ``external legs'' (solid lines) are real. In the laboratory they
show up as data, e.g. particle tracks. You see them in an experiment. The
dotted line is internal. That is a ``virtual'' particle. Bohr has warned us
not ask about the virtual quantum process. You are not meant to know (dear
reader) about the virtual. Do not ask! For example, there are two
(Bohr) complimentary views: (i) four momentum space amplitudes
and (ii) space-time amplitudes.

(i) In momentum space the ``virtual particle'' dotted line in Fig.1 has a
nice four momentum label $P_{virtual}$. That seems simple enough. The only
problem is that
$$
[P_{virtual}^2+(Mc)^2]\ne 0. \ \ (possibly \ off\ mass\ shell). \eqno(1a)
$$
Virtual particles may not be on the ``mass shell''. Real particles are on
the mass shell. Virtual particles are not real. Bohr warned us not to ask.

(ii) The space-time ``virtual particle'' in the words of Feynman ``...does
anything it likes...''. It goes forward in time. It goes backward in time.
It goes space-like. Any old speed said Feynman. Anything it likes,
$$
x_{virtual}^2=r_{virtual}^2-c^2t_{virtual}^2>0\ \ (possibly\ superluminal\
 speed). \eqno(1b)
$$
STOP IT! Bohr dictates the if it is virtual, then do not ask. Virtual
particles are not real. There will be no data.

(iii) But experimentalists tell us every day that there are data on virtual
particles you do not actually see. Look again at the Feynman diagram of
Fig.1. Suppose the external legs are measured tracks shown in postscript
colors on an experimentalist's transparency (right after we see a small
person standing next to a huge machine). {\it We do not see the dotted
line}. The right tracks lead to a primary vertex and left tracks emerge from
the secondary vertex. The dotted line is a particle that moved from the
primary vertex to the secondary vertex. {\it Only nobody sees it!} It had
no charge! It left no track! But we all know it was there. This, by our
definition, is an {\it experimental virtual particle}. It moved from vertex
to vertex and nobody saw it. We are not talking about the moon. We are
talking about an elementary particle. If you do not see it, then it is
virtual. But the experimentalists tell us that this virtual particle moved
from the primary vertex $y$ to the secondary vertex $x$ along a classical
path at constant (four velocity) $v$ and on the mass shell,
$$
x^\mu =y^\mu +v^\mu \tau, \ \ (experimental\ virtual\ particle ), \eqno(2a)
$$
$$
P_{virtual}^2 +(Mc)^2=M^2(v^2+c^2)=0,\ \ (experimental\ virtual\ particle ).
\eqno(2b)
$$
Comparing the experimentally employed Eqs.(2) to the exciting theoretical
possibilities in Eqs.(1), we find that the experimentalist's virtual particle
is rather dull and classical, i.e. {\it it looks real (not virtual) even if
you do not see it}.

Bohr had all the answers. He threw in the correspondence principle along
with the complimentarity of momentum space and space-time. Only Einstein was
worried. The correspondence principle says that when the action of the
particle is large on the scale of $\hbar $ the particle is classical {\it
even if you do not look at it}. The particle going from the primary to the
secondary vertex had to move so far that the action was very large compared
with $\hbar $. The particle may do ``anything it likes'', but when the
action is large it ``likes'' to be classical with overwhelming probability.
To do otherwise would have such small probability as to be a miracle.
Miracles can happen, but they are hard to duplicate. There are no good
statistics on miracles.

The mathematics is that the propagator from vertex to vertex
$$
[-\partial _\mu \partial ^\mu +\kappa ^2]D(x-y)=\delta (x-y),\ \ \ \kappa
=(Mc/\hbar ),\eqno(3)
$$
has the Dirac-Feynman path integral representation
$$
D(x-y)=\int_{z(0)=y}^{z(\tau )=x}\prod_\tau \left( {\frac{{\cal D}p{\cal D}z%
}{2\pi \hbar }}\right) exp{\frac i\hbar }\int_y^x[p_\mu dz^\mu -{\cal H}%
(p)d\tau ],\eqno(4)
$$
where
$$
{\cal H}(p)={\frac 1{2M}}[p^2+(Mc)^2].\eqno(5)
$$
If the particle were classical, then the free particle Hamilton-Jacobi
equation \cite{s4}
$$
-{\frac{\partial S}{\partial \tau }}={\cal H}\left( p={\frac{\partial S}{%
\partial x}}\right) ,\eqno(6a)
$$
$$
S(x-y,\tau )=-{\frac{Mc^2}2}\left( \tau -{\frac{(x-y)^2}{c^2\tau }}\right) ,%
\eqno(6b)
$$
would yield the classical experimental Eqs.(2) via the equations
$$
-{\frac{\partial S}{\partial \tau }}=0,\ \ \ S=-Mc\sqrt{-(x-y)^2},\ \ \ M{%
\frac{dx^\mu }{d\tau }}=\partial ^\mu S.\eqno(6c)
$$
The actual space-time propagator in the Schwinger ``proper time''
representation is given by
$$
D(x-y)={\frac M{8\pi ^2\hbar }}\int_0^\infty \left( {\frac{d\tau }{\tau ^2}}%
\right) exp{\frac i\hbar }S(x-y,\tau ),\eqno(7)
$$
where the proper time $\tau $ internal to the particle need not be the
laboratory proper time $\sqrt{-(x-y)^2/c^2}$ between the vertex events.
The particle has all the proper time in the world to do what it likes;
forward in time, backward in time, spacelike ... and so forth. But the Bohr
correspondence principle tells us that when the action in Eq.(7) obeys
$|S|>>\hbar $, then the stationary phase evaluation of the proper-time
integral in Eq.(7) yields $\tau =\sqrt{-(x-y)^2/c^2}$ and what the particle
then likes to do is to be classical. The experimentalist has every right
to suppose that the ``virtual particle'' is classical even if nobody detects
the path, and likewise for the moon. If Bohr's correspondence holds true,
then all is right with this boring world. But when a particle shows up that
exhibits virtual quantum interference, in {\it spite} of the correspondence
principle, everybody says that this particle must be very strange.

\section{The ``$K^o\overline{K}^o$'' Particle}

The ``$K^o\bar{K}^o$'' virtual particle is not merely strange, it has some
deep psychological problems. It cannot decide whether to exhibit itself as a
$K^o$ going forward in time or as $\bar{K}^o$ going backward in time. (The
language is that of St\"uckelberg and Feynman.) If the dotted line in Fig.1
is ``$K^o\bar{K}^o$'' as a virtual particle, it will never be classical even
for $|S|>>\hbar$ because {\it any decent free classical particle} (or
classical anti-particle) will make up its mind once and for all which
direction in time it wants to go and then it learns to live in psychological
peace with its decision. Hence, the Bohr correspondence principle works very
well except for those cases where it does {\it not} work very well.

If a ``$K^o\bar{K}^o$'' T violating particle (we assume TCP=1) goes from a
primary to a secondary vertex its inner turmoil refuses the classical limit
of the Bohr correspondence principle and remains truly virtual, oscillating
with quantum interference phase factors. Very good. Now the Bohr dictate
is that we cannot ask what the ``$K^o\bar{K}^o$'' is doing. But it is really
hard not to try and form a mental picture such as maybe the particle path is
still classical but ``wobbles'' a little bit.

The mathematics is that the (two by two) non-Hermitian mass matrix of the
``$K^o\bar K^o$'' \cite{s5,s6}
$$
{\cal M}=M-i(\hbar \Gamma /2c^2),\eqno(8)
$$
produces the (two by two) ``wobbly'' propagator
$$
{\cal D}(x-y)={\frac{{\cal M}}{8\pi ^2\hbar }}\int_0^\infty \left( {\frac{%
d\tau }{\tau ^2}}\right) exp{\frac{i{\cal M}}{2\hbar }}[-c^2\tau +{\frac{%
(x-y)^2}\tau }].\eqno(9)
$$
The Bohr correspondence principle tells us to evaluate the proper time
integral by stationary phase yielding the particle proper time as the
laboratory proper-time so that asymptotically the particle has the
laboratory proper-time
$$
{\cal D}(x-y)\sim e^{-i{\cal M}c^2\tau /\hbar },\ \ \ -c^2\tau ^2\approx
(x-y)^2,\eqno(10)
$$
and we are allowed (if we choose) to think of the propagator Eqs.(9) and
(10) as a proper time Schr\"odinger equation
$$
i\hbar {\frac \partial {\partial \tau }}\left(
\begin{array}{c}
A_{K^o}(\tau ) \\
A_{\overline{K}^o}(\tau )
\end{array}
\right) ={\cal M}c^2\left(
\begin{array}{c}
A_{K^o}(\tau ) \\
A_{\overline{K}^o}(\tau )
\end{array}
\right) ,\eqno(11)
$$
but we are not allowed to think of the ``$K^o\bar K^o$'' traveling from a
primary vertex to a secondary vertex as real (i.e. classical). It is truly
an experimental virtual particle. Typical of this kind of experiment is the
CPLEAR project \cite{s7} where the Feynman diagram of a typical event is shown
in
Fig.2. The ``$K^o\bar K^o$'' is not observed (there is no track) and it is
truly virtual because you cannot Bohr correspond when you need a
Schr\"odinger equation. Bohr has warned us about getting data from a
quantum object.

One notes that the four velocity of the ``$K^o\bar{K}^o$'' particle
$$
v^\mu ={\frac{(x^\mu -y^\mu)}{\tau}},\ \ \ \tau=\sqrt{\frac{-(x-y)^2}{c^2}},
\eqno(12)
$$
is determined if we know the space-time positions of the vertex events in
which the classical paths (legs of the Feynman diagrams) are located. That
the ``$K^o\bar{K}^o$'' particle is still virtual is due to the fact that it
has two possible (``long'' or ``short'') four momenta
$$
p_L=M_Lv\ \ or\ \ p_S=M_Sv, \eqno(13)
$$
so that a typical amplitude
$$
A(\tau )=A_L e^{-\Gamma_L \tau /2}e^{-iM_Lc^2\tau /\hbar} +A_S e^{-\Gamma_S
\tau /2}e^{-iM_Sc^2\tau /\hbar}, \eqno(14)
$$
has two ``plane wave'' fixed momentum states going from vertex $y$ to vertex
$x$, e.g. in Eqs.(12), (13) and (14)
$$
exp(-iM_jc^2\tau /\hbar)=exp(ip_j\cdot(x-y)/\hbar),\ \ j=L\ or\ S. \eqno(15)
$$
The notion that the single virtual ``$K^o\bar{K}^o$'' has no trouble making
up its mind as to its velocity but becomes confused about what should be its
momentum, due to its mass splitting $\Delta M=(M_L-M_S)$, is crucial to the
conventional analysis of single ${\cal D}(x,y)$ propagator experiments (even
when a regenerator is included as part of the propagator). While
theoretically this seems all right, e.g. the value of $\Delta M$ found in
CPLEAR should agree with $\Delta M$ in regenerator experiments \cite{s8},
we still feel somewhat uncomfortable with the result.

If the single virtual (large spreading wave packet) ``$K^o\bar{K}^o$'' is in
a superposition of two plane wave momentum states, as in Eqs.(14) and (15),
and this particle scatters off other particles, should not some other
particles also have trouble deciding in which momentum states {\it they are}?
After all, no matter what the state of mind of each particle may be, there
should be four momentum conservation. In other words, if an initial state is
in a superposition of two different incoming four momenta ($P_1$ and $P_2$),
i.e.
$$
|\Psi >=c_1|in,P_1>+c_2|in,P_2>, \eqno(16a)
$$
and if the scattering operator conserves four momenta, i.e.
$$
S|\Psi >=c_1|out,P_1>+c_2|out,P_2>, \eqno(16a)
$$
then shouldn't the outgoing particles (with the same amplitudes $c_1$ and
$c_2$) also be in superposition? The confused particles would also be
virtual, since Bohr dictates that those particles which leave tracks are
real and classical and thereby do not maintain ``superpositions''. The
superposition principle has to be relegated to the virtual processes that
you do not observe, and don't ask what a virtual particle is doing because
there will be no data!

Let us now proceed to ``two propagator'' experiments wherein there are at
least two virtual particles. These include the proposed $\Phi $ factory
experiments, and here momentum conservation has some unusual
implications.

\section{Two Propagator Experiments}

Shown in Fig.3 (below) are typical Feynman diagrams of a previous two
propagator experiment \cite{s9} (which involves a virtual $\Lambda $)
as well as a
virtual ``$K^o\bar{K}^o$''. Also shown is a typical two propagator Feynman
diagram for a process anticipated at the $\Phi $ factory \cite{s10}. If the
``$K^o\bar{K}^o$'' internal propagators are forced to have a superposition
of two four momentum states, then we have predicted that even the
$\Lambda $ would be forced to be in a superposition of two momentum
states. ``$\Lambda $ oscillations" \cite{s11} would be the result of such a
superposition, born by a combination of the mass splitting
$\Delta M =(M_L-M_S)$ and of momentum conservation at the
${\bar K^o}+p^+\rightarrow \Lambda +\pi^+ $ vertex. The inner turmoil of the
``$K^o\bar{K}^o$'' spreads to any other virtual particle connected to it
in the Feynman diagram.

To see that a discussion of the real and the virtual is not trivial,
consider the $\Phi $ factory two virtual K meson process. The total
four momentum is determined by the real electron positron pair
producing the $\Phi $,
$$
P_{\Phi }=p_{e^+}+p_{e^-}=
p_{L,left}+p_{S,right}=p_{S,left}+p_{L,right}. \eqno(17)
$$
Now, from the two propagators you try to put together a simple two
particle wave function (not quite a Schr\"odinger wavefunction because
there are too many ``times''),
$$
\Psi_{simple}(\tau_{right},\tau_{left})=
A_L(\tau_{right})A_S(\tau_{left})-
A_L(\tau_{left})A_S(\tau_{right}), \eqno(18)
$$
where
$$
A_j(\tau )=e^{-\Gamma_j\tau /2}e^{-iM_jc^2\tau /\hbar}
\ \ \ (j=L\ or S). \eqno(19)
$$
But we find that  the total momentum conservation Eq.(17)
cannot be maintained by Eq.(18); e.g., see Eq.(15),
$$
A_L(\tau_{right})A_S(\tau_{left})\sim
exp{i\over \hbar}(p_L\cdot x_{right}+p_S\cdot x_{left}),
\eqno(20a)
$$
$$
A_L(\tau_{left})A_S(\tau_{right})\sim
exp{i\over \hbar}(p_L\cdot x_{left}+p_S\cdot x_{right}),
\eqno(20b)
$$
but the two plane waves in $\Psi_{simple}$ will not conserve
total momentum unless $p_L=p_R$ ({\it false!}). This ultimately violates
the one velocity and two momenta inner turmoil Eq.(13), and would thereby
afford an undue psychological relief to the mass difference induced
$K^o\bar{K}^o$ oscillations. So, the simple wave function in Eq.(18) appears
inadequate for describing the two propagator result, if conservation
of total momentum is strictly applied \cite{s12}.

What is needed for conservation of total momentum is precisely what is
written in Eq.(17), which {\it actually describes conservation of
momentum}. Only one needs more proper times in the amplitude, because you
need more momenta in the plane waves
$$
\Psi(\tau_{L,right},\tau_{L,left},\tau_{S,right},\tau_{S,left})=
$$
$$
A_L(\tau_{L,right})A_S(\tau_{S,left})-
A_L(\tau_{L,left})A_S(\tau_{S,right}). \eqno(21)
$$
What can we do with all of those times (don't ask!)?  Here are the
numbers we promised you (dear reader) in Sec.1. Actually it is one
(singular) number but it is predicted to occur very often in the data.

When you absolute square amplitudes with a wave function that has
two terms, e.g. Eq.(21), the cross terms give you interference. For the
problem at hand the interference phase is given by
$$
\theta ={c^2\over \hbar}[M_L(\tau_{L,left}-\tau_{L,right})-
M_S(\tau_{S,left}-\tau_{S,right})]. \eqno(22)
$$
Defining averages and differences
$$
\tau_i=(1/2)(\tau_{L,i}+\tau_{S,i}),\ \ (i=left\ or\ right), \eqno(23a)
$$
$$
\Delta \tau_i=\tau_{L,i}-\tau_{S,i},\ \  (i=left\ or\ right), \eqno(23b)
$$
$$
\bar{M}=(1/2)(M_L+M_S),\ \ \Delta M=(M_L-M_S), \eqno(23c)
$$
we find from Eqs.(22) and (23),
$$
(\hbar \theta /c^2) =\bar{M}(\Delta \tau_{left}-\Delta \tau_{right})
+\Delta M(\tau_{left}-\tau_{right}). \eqno(24)
$$
The first term on the left hand side Eq.(24) would not be present if we
employed Eq.(18). But again by conservation of momentum we find the
two terms on the right hand side of Eq.(24) approximately equal, and
thus the factor of two in our final result
$$
\theta = 2(c^2 \Delta M/\hbar)(\tau_{left}-\tau_{right}). \eqno(25)
$$
The factor of ``$2$'' in Eq.(25) is the central result of this work.

\section{Conclusion}

We are {\it emotional} about a factor of ``2'' because we believe it says
something unexpected about what is virtual and what is real. Others
are {\it very emotional} about it because they calculate the factor
to be ``$1$''!. But the nice thing about it, is that the answer will
surely be found at the $\Phi $ factory, with the experimentalists as
the final referees. In Italy, where Galileo taught us that experimental
data are the final arbiters in a theoretical dispute, what else could
happen?

\end{document}